\documentclass{PoS}
\usepackage{wrapfig}
\usepackage{subfigure}

\usepackage{graphicx}
\usepackage{dcolumn}
\usepackage{bm}
\usepackage{epsf}
\usepackage{amssymb,amsmath,amsfonts,amsthm, amscd, mathrsfs,helvet}

\usepackage{verbatim}
\usepackage{psfrag}
\usepackage[all]{xy}
\usepackage{cite}
\usepackage{epic}

\newcommand{\beq}{\begin{equation}}
\newcommand{\eeq}{\end{equation}}
\newcommand\beqa{\begin{eqnarray}}
\newcommand\eeqa{\end{eqnarray}}
\newcommand\bea{\begin{array}}
\newcommand\eea{\end{array}}
\newcommand\ba{\begin{array}}
\newcommand\ea{\end{array}}
\newcommand{\nn}{\nonumber}

\newcommand{\neqa}{\nonumber\end{eqnarray}}
\newcommand{\la}{\label}

\newcommand{\eq}[1]{eq.(\ref{#1})}

\newcommand{\hf}{\frac{1}{2}}

\renewcommand{\d}{\partial}
\renewcommand{\O}{{\cal O}}

\newcommand{\<}{{\langle}}
\renewcommand{\>}{{\rangle}}

\def\l{\lambda}

\def\s{\sigma}
\def\a{\alpha}
\def\b{\beta}
\def\th{\theta}

\def\({\left(}
\def\){\right)}
\def\[{\left[}
\def\]{\right]}

\def\<{\langle}
\def\>{\rangle}

\newcommand{\re}{\relax{\rm I\kern-.18em R}}

\newcommand{\stp}{\tilde p\hspace{-.40em}/}

\def\a{{\alpha}}

\def\hf{\frac{1}{2}}

\def\({\left(}
\def\){\right)}
\def\[{\left[}
\def\]{\right]}

\def\su2{{SU(2)}}

\def\tr{{\rm tr}}

\def\sG{\,\slash\!\!\!\! G}

\def\pint{-\hskip-0.41cm \int}

\title{\begin{flushright}
\vspace{-2.9cm}
\small \textmd{LPTENS-07/13\\
hep-th/0703137}
\vspace{2.9cm}
\end{flushright}
Classical limit of Quantum Sigma-Models from Bethe
Ansatz}
\ShortTitle{Classical limit of Quantum Sigma-Models from
Bethe Ansatz}

\author{Nikolay~Gromov,$^{ab}$ \speaker{Vladimir~Kazakov}$^{a}$ and Pedro~Vieira$^{ac}$\\
\llap{$^a$}Laboratoire de Physique Th\'eorique
de l'Ecole Normale Sup\'erieure\footnote{Unit\'e mixte du C.N.R.S.
et de l' Ecole Normale Sup\'erieure, UMR 8549.},
24 rue Lhomond, Paris, CEDEX 75231, France; l'Universit\'e Paris-VI\\
\llap{$^b$}St.Petersburg INP,
 Gatchina, 188 300, St.Petersburg, Russia\\
\llap{$^c$}Departamento de F\'\i sica e Centro de F\'\i sica do
Porto, Faculdade de Ci\^encias da Universidade do Porto, Rua do
Campo Alegre, 687, \,4169-007 Porto, Portugal}

\abstract{In these proceedings we review the results of \cite{Kazakov:2004qf,GKSV,Gromov:2006cq}. We show on the example of the $SU(2)$ chiral-field how to reproduce
the classical finite gap
   solutions for a large class  of integrable sigma models from their exact quantum solutions.
   These solutions are usually formulated as  Bethe ansatz equations   for physical particles on
    a circle, with the interaction  given by the  factorized S-matrix conjectured from
    Zamolodchikovs' bootstrap procedure.  Our method opens a new systematic way to justify
    this procedure.
      As an application of our method to the integrability in AdS/CFT correspondence,
      we reproduce the asymptotic string Bethe ansatz conjectured eartlier in the
      $S^3\times R_t$ sector of the
  Green--Schwarz--Metsaev--Tseytlin superstring. The role of the Virasoro constraints
  in this setup is clarified.}

\FullConference{Solvay workshop "Bethe Ansatz: 75 Years Later"\\ Brussels, Belgium\\ October 19-21, 2006}

\begin{document}

\section{Introduction}

The classical integrable\footnote{i.e. having an infinite number of
integrals of motion} two-dimensional non-linear sigma models are
relatively easy to solve. At least, when the corresponding Lax pair
is known, one can construct a large class of the so called classical finite
gap solutions \cite{Novikov:1984id}. These solutions are known to
constitute a dense (in the sense of parameters of initial
conditions) subset in the  space of solutions of the model.

However, the   quantization of such classically integrable
sigma-models usually creates substantial problems and is known to be
virtually impossible to do in the direct way, in terms of the
original degrees of freedom of the classical action. The existing
quantum solutions  are usually based on plausible assumptions which
are difficult to prove in a systematic way.

There were a few successful, though not completely justified,
attempts to find the quantum solutions of $SU(N)\times SU(N)$
principal chiral field model (PCF),  starting from the original
action. A.~Zamolodchikov and Al.~Zamolodchikov
\cite{Zamolodchikov:1977nu} found the factorizable bootstrap
S-matrices for the $O(N)$ sigma models, later generalized to many
other sigma models. The $O(4)$ case which we are focused on in this
paper, is equivalent to the $SU(2)\times SU(2)$ PCF.
 Polyakov and Wiegmann
\cite{PolyakovWiegmann,Wiegmann:1984mk} found the equivalent
non-relativistic integrable Thirring model reducible in a special
limit to the PCF. Faddeev and Reshetikhin \cite{Faddeev:1985qu}
proposed the "equivalent" double spin chain for the $SU(2)\times
SU(2)$ PCF. In both cases, the equivalence is based on subtle
assumptions, difficult to verify, though both approaches perfectly
reproduce the solution following from the S-matrix approach
\cite{Wiegmann:1984ec}.

The verification of such solutions is usually based on the
perturbation theory, large $N$ limit or Monte-Carlo simulations
\cite{Zamolodchikov:1977nu,Wiegmann:1984ec,FKW,Niedermayer}.

Here we address this question in a more systematic way. Namely, we
will reproduce all classical finite gap solutions of a sigma model
from the Bethe ansatz solution for a system of physical particles on
the space circle, in a special large density and large energy limit.
We shall call it the continuous limit though, as we show, it is the
actual classical limit of the theory. We will see that in this limit
the Bethe Ansartz equations (BAE) diagonalizing the periodicity
condition, will be reduced to a Riemann-Hilbert problem. Such a
limit in the Bethe ansatz equation is similar to the one considered
in
\cite{ReshetikhinSmirnov,Beisert:2003xu,Sutherland,Korchemsky:1997ve})
defining the algebraic curve of the finite gap method for the
underlying classical model.

We demonstrate the method inspired by \cite{Mann:2005ab} and worked
out in \cite{GKSV,Gromov:2006cq} for the $SU(2)\times SU(2)$
principal chiral field (PCF) with the action\footnote{note that the
coupling $\lambda$ is chosen here as the 'tHooft coupling in the
AdS/CFT correspondence context.}
 \begin{equation}
S=\frac{\sqrt{\l}}{8\pi} \int d\s d\tau\,  \tr\,\d_a g^\dag\d_a g,\qquad g\in SU(2) \,.
\la{ACTIONSU2}
\end{equation}
In \cite{GKSV} we also repeated this construction for the $O(6)$
sigma-model and explained how the generalization to the $O(2n)$
model can be done in a trivial way. In fact, as it will be clear
below, the method seems to be general enough to work for all
sigma-models described by a factorizable  bootstrap S-matrix. Hence
it gives a new way to relate, in a general and systematic way, the
classical and quantum integrability.


The model (\ref{ACTIONSU2}) is equivelent to  the $O(4)$ sigma model
where the fundamental field is the four dimensional unit vector
$\vec{X}(\sigma,\tau)$. Therefore, at least classically, it can be
used to study a string on the  $S^3\times R_1$ background. Indeed,
our main motivation for this study was the search for new approaches
in the quantization of the Green--Schwarts--Metsaev--Tseytlin
superstring on the $AdS_5\times S^5$ which is  classically (and
most-likely quantum-mechanically as well) an integrable field
theory. The simplest nontrivial subsector  of it is described by the
sigma model on the subspace $S^3\times R_t$, where $R_t$ is the
coordinate corresponding to the AdS time. The time direction will be
almost completely  decoupled from the dynamics of the rest of the
string coordinates, appearing only through the Virasoro conditions.
These conditions are a selection rule for the states of the theory
or, better to say, for the classical solutions appearing when we
pick the classical limit in Bethe equations. The degrees of freedom
eliminated in this way are the longitudinal modes associated with
the reparametrization invariance of the string.

Of course, in the absence of the fermions and of the AdS part of the
full 10d superstring theory, this model will be asymptotically free
and will not be suitable as a viable (conformal) quantum string
theory. Nevertheless, in the classical limit we shall encounter the
full  finite gap solution of the string in the  $SO(4)$ sector found
in \cite{Kazakov:2004qf}. The method can be generalized  to the
$SO(6)$ sector in \cite{Beisert:2004ag} and  hopefully to the full
Green--Schwarts--Metsaev--Tseytlin superstring on the $AdS_5\times
S^5$ space, including fermions, where the finite gap solution was
constructed in \cite{Beisert:2004ag} (although it appears to be more
difficult for the last, and the most interesting, system).

At the end of the paper we go slightly further and derive from these
BAE the conjectured asymptotic string Bethe ansatz (the so called
AFS-equation \cite{Arutyunov:2004vx}) with its nontrivial dressing
factor to the leading order in large $\lambda$ which is known to
capture some quantum effects, such as level spacing \cite{GV}.


\subsection{Classical  $SU(2)\times SU(2)$ Principal Chiral Field}

In this section we will review the classical finite gap solution of
the $SU(2)\times SU(2)$ principal chiral field. We will essentially
go through the construction of \cite{Kazakov:2004qf}\footnote{with a
little generalization to the excitations of both left and right
sectors} to fix the notations for the easy comparison with the
quantum Bethe ansatz solution of the model. As mentioned in the
introduction, classically this model can be used to describe the
string on $S^3\times R_t \subset AdS_5\times S^5$. At the quantum
level, even dropping all the rest of the degrees of freedom, one
might still expect to capture some features of the full superstring
theory. As we will see in the latter sections, this is indeed the
case.

\subsubsection{The model}

The action (\ref{ACTIONSU2}) possesses the obvious global symmetry
under the right and left multiplication by $SU(2)$ group element.
The currents associated with this symmetry are, respectively,
\beq  j^R \equiv j = g^{-1}d g\,,\qquad j^L = d g g^{-1}\,,
\eeq
and the corresponding Noether charges read
\beq Q_R=\frac{i\sqrt{\lambda}}{4\pi}\int_0^{2\pi}
d\sigma\ \tr\(j^R_\tau\, \tau^3\),\;\;\;\;\;Q_L=\frac{i \sqrt{\lambda}}{4\pi}\int_0^{2\pi} d\sigma\
\tr\( j^L_\tau \tau^3\)\,.\;\;\;\;\; \la{Charges} \eeq
In the quantum theory these charges are positive integers\footnote{It will
be important for future comparisons to notice that the normalization
of the generators is such that the smallest possible charge is $1$ as
follows from the Poisson brackets for the current.}.

Virasoro conditions read $\tr \(j_\tau\pm
j_\sigma\)^2=-2\,\kappa_\pm^2$, where we used the residual
reparametrization symmetry to fix the $AdS$ global time $Y$ to \beq
Y=\frac{\kappa_+}{2}(\tau+\sigma)+\frac{\kappa_-}{2}(\tau-\sigma)\,.\la{Y}
\eeq

Finally, from the action, we read off the energy and momentum as
\beq \la{EPclassic} E^{\,\rm cl}\pm P^{\,\rm
cl}=-\frac{\sqrt{\lambda}}{8\pi}\int_0^{2\pi}\tr \(j_\tau\pm j_\sigma
\)^2d\sigma=\frac{\sqrt{\lambda}}{2}\,\kappa^2_\pm\,. \eeq

\subsection{\la{AnPrSO6}Classical Integrability and Finite Gap Solution}\label{first}

The equations of motion and the fact that the current is of the form $j=g^{-1}dg$
can be encoded into a single flatness condition for a Lax connection over
the world-sheet\cite{Novikov:1984id},
\beq \[ \partial_\sigma - \frac{x\, j_\tau+j_\sigma}{x^2-1}, \partial_\tau
- \frac{x\, j_\sigma+j_\tau}{x^2-1}\]=0. \label{LAX} \eeq
In particular, we can then use this flat connection to define the monodromy matrix
\beq\label{MONO}
\Omega(x)=\stackrel{\leftarrow}{P}\exp\int_0^{2\pi}d\sigma\,
\frac{x j_\tau+j_\sigma}{x^2-1} . \eeq
By construction $\Omega(x)$ is a unimodular matrix (and also unitary for real $x$)
whose eigenvalues can therefore be written as
\beq  \left(e^{i\tilde p(x)},e^{-i\tilde p(x)}\right)  \eeq
where $\tilde p(x)$ is called the quasi-momentum. These \textit{functions of $x$}
do not depend on time
$\tau$ due to (\ref{LAX}) and provide therefore an infinite set of classical
integrals of motion of the model.

From the explicit expression (\ref{MONO}) we can determine the
behaviour of the quasi-momentum close to $x=\pm1,0,\infty$. Using
(\ref{EPclassic}) and (\ref{Charges}), we obtain
\beqa \la{LCH}    \tilde p(x)&\simeq&  -\frac{\pi\kappa_\pm}{x\mp 1} \,, \\
\la{PXZERO}   \tilde p(x) &\simeq& 2\pi m+{2\pi Q_L\over
\sqrt{\lambda}}x \,,  \\
\la{GLCH} \tilde p(x)&\simeq&-\frac{2\pi Q_R}{\sqrt{\lambda}\,
}\frac{1}{x} \,.   \eeqa
Since, by construction, $\Omega(x)$ is analytical in the whole plane
except at $x=\pm 1$ where it develops essential singularities, it follows from
\eq{OMEGAP} that for $x\neq \pm1$ the only
singularities of
\beq\la{OMEGAP}  \tilde  p\,'(x) =-\frac{1}{\sqrt{4-\(\tr\,\Omega(x)\)^2}}
\frac{d}{dx}\tr\,\Omega(x) \,.
\eeq
are of the form
\beq \tilde p\,'\,(x\to x_k)\simeq
\frac{1}{\sqrt{x-x_k}}   \,.\eeq
\begin{figure}[t]
\centering
        \includegraphics{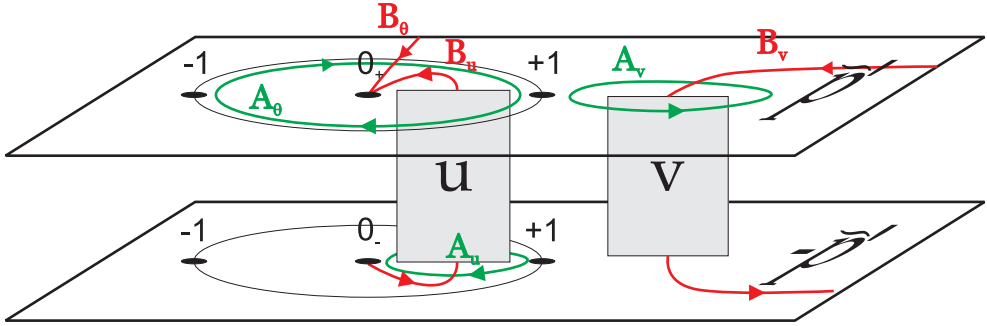}
    \caption{\textit{Algebraic curve from the finite gap method. $u$ and $v$ cuts correspond
    to cuts inside and outside the unit circle respectively.}}
\end{figure}

If we are looking  for a finite gap solution the number $K$ of these
cuts is finite and we conclude that $\tilde p'(x),-\tilde p'(x)$ are
two branches  of an analytical function defined by a hyperelliptic
curve (see fig.1),
\beq
(p')^2=\frac{P^2(x)}{Q(x)}, \label{pPOL}
\eeq
where $Q(x)$ has $2K$ zeros and the order of $P(x)$ is fixed by the
large $x$ asymptotics \eq{GLCH}. We denote the branch cuts of $p'(x)$
by $u$ ($v$) cuts if they are inside (outside) the unit circle. These
cuts are the loci where the eigenvalues of the monodromy matrix become
degenerate. Thus, when crossing such cut the quasi-momentum may at most
jump by a multiple of $2\pi$ which characterizes each cut,
\beq
\stp(x)= \pi n_k,\qquad x\in {\cal C}_k \la{ps} \eeq
where $\stp(x)$ is the average of the quasi-momentum above and below the cut, %
\beq\la{SLASHP} \stp(x) \equiv \hf \(\tilde p(x+i0)+\tilde p(x-i0)\)\,.\eeq

Each cut is also parameterized by the filling fraction numbers which we define as integrals along
$A$-cycles of the curve (see fig.1) \footnote{It was pointed out in
\cite{Beisert:2004ag,Beisert:2005di} and shown in
\cite{Dorey:2006zj} that $S_i^{u,v}$ are the action variables so
that quasi-classically they indeed become integers. We will also
find a striking evidence for this quantization on the string side
when finding the classics from the quantum Bethe ansatz where these
quantities are naturally quantized. Indeed, from the AdS/CFT
correspondence these filling fractions are expected to be integers
since this is obvious on the SYM side
\cite{Kazakov:2004qf,Beisert:2005di}.  }
\beqa\la{ACYCLE} S_i^v=-\frac{\sqrt{\lambda}}{8\pi^2
i}\oint_{A^v_i}\tilde p(x)\(1-\frac{1}{x^2}\)dx,\;\;\;\;\;
S_i^u=\frac{\sqrt{\lambda}}{8\pi^2 i}\oint_{A^u_i}\tilde
p(x)\(1-\frac{1}{x^2}\)dx \,.
\eeqa
Finally, imposing
(\ref{ps},\ref{ACYCLE},\ref{LCH},\ref{PXZERO},\ref{GLCH}) one fixes
completely the undetermined constants in (\ref{pPOL}).

\section{Quantum Bethe Ansatz and Classical Limit: $O(4)$ Sigma-Model \label{sec3}}

We will describe a  quantum state of the $O(4)$ sigma model by a system of $L$ relativistic particles of
mass $\mu/2\pi$
 put on a circle of the length $2 \pi$. The momentum and the energy of each particle can be suitably parametrized by its rapidity as $p=\frac{\mu}{2\pi} \sinh\theta$ and $e=\frac{\mu}{2\pi} \cosh\theta$ so that the total energy and momentum will be given by
 \begin{eqnarray}
\la{P_SSL} P&=&\frac{\mu}{2\pi}\sum\limits_{\a=1}^L\sinh(\pi\theta_\alpha) \,, \\
\la{E_SSL} E&=&\frac{\mu}{2\pi}\sum\limits_{\a=1}^L\cosh(\pi\theta_\alpha) \,.
\end{eqnarray}
  These
particles transform in the vector representation under $O(4)$
symmetry group or in the bi-fundamental representations of
$SU(2)_R\times SU(2)_L$. The scattering of the particles in this
theory is known to be elastic and factorizable: the relativistic
S-matrix $\hat S\(\th_1-\th_2\)$  depends only on the difference of
 rapidities of scattering particles $\th_1$ and $\th_2$ and obeys the
Yang--Baxter equations. As was shown in \cite{Zamolodchikov:1977nu}
(and in \cite{Wiegmann:1984mk,Wiegmann:1984ec,Wiegmann:1984pw,Ogievetsky:1984pv} for the general principle chiral field) these
properties, together with the unitarity and crossing-invariance,
define essentially unambiguously the S-matrix $\hat S$. Let us recall briefly how the bootstrap program goes. From the symmetry of the problem we know that
\beq \hat S=\hat S_L\times\hat S_R \eeq
where $S_{L,R}$ are built by use of the two $SU(2)$ invariant tensors and can therefore be written as
\begin{eqnarray*}
S_{R,L}(\theta)^{a'b'}_{a\,b}=\frac{S_0(\theta)}{\theta-i}\left(\theta\,
\delta^{a'}_{a}\delta^{b'}_{b}-i\,f(\theta)\,\delta^{b'}_{a}\delta^{a'}_{b}\right)\,.
\end{eqnarray*}
Imposing the Yang-Baxter equation on $\hat S$ yields $ f(\theta)=1
$, while the unitarity constrains the remaining unknown function to
obey \beq S_0(\theta)S_0(-\theta)=1\, \la{unit} \eeq and crossing
symmetry requires \beq S_0(\theta)=\(1-\frac{i}{\theta}\)
S_0(i-\theta) \,. \la{crossing} \eeq From (\ref{unit}),
(\ref{crossing}) and the absence of poles on the physical strip
$0<\theta<2$ one can compute the scalar factor:
$S_0(\theta)=\frac{\Gamma\left(-\frac{\theta}{2i}\right)\Gamma\left(\frac{1}{2}
+\frac{\theta}{2i}\right)}{\Gamma\left(\frac{\theta}{2i}\right)
\Gamma\left(\frac{1}{2}-\frac{\theta}{2i}\right)}$.
For our purpose we just need
the much easier to extract large $\theta$ asymptotics,
 \beq\label{ASSS}  i\log S^2_0(\theta)\sim 1/\theta + O(1/\theta^3) \,.  \eeq

\subsection{Bethe Equations for Particles on a Circle}

 When this system of particles is put into a
finite 1-dimensional periodic box of the length $\cal L$ the set of
rapidities of the particles $\{\theta_\a\}$ is constrained by the
condition of  periodicity of the wave function $|\psi\rangle$ of the
system,
\begin{eqnarray}
|\psi\rangle=e^{i\mu
\sinh\pi\theta_\a}\overleftarrow{\prod_{1}^{\a-1}}\ \hat
S\(\th_\a-\th_\b\) \overrightarrow{\prod_{N}^{\a+1}}\ \hat
S\(\th_\a-\th_\b\) |\psi\rangle \label{MBAE}
\end{eqnarray}
where the first term is due to the free phase of the particle and
the second is the product of the scattering phases with the other
particles. The arrows stand for ordering of the terms in the
product and $\mu=m_0{\cal L}$ is a  dimensionless parameter.
Diagonalization of both the L and R factors in the process of fixing
the periodicity (\ref{MBAE}) leads  to the following set of Bethe
equations \cite{Zamolodchikov:1992zr} which may be found from \eq{MBAE} by the algebraic Bethe
ansatz method  \cite{Faddeev:1979gh,Kulish:1981gi} \footnote{We took
the logarithms of the Bethe ansatz equations in their standard,
product form. This leads to the integers $m_\alpha,n_j^{u},n_j^{v}$
defining the choice of the branch of logarithms.}
\begin{eqnarray}
2\pi m_\alpha&=&\mu \sinh\pi\theta_\a- \sum_{\beta\neq \alpha}^L\,i\log
S_0^{\,2}\(\th_\a-\th_\b\) \nn \\&-&
\sum_j^{J_u}i\log\frac{\th_\a-u_j+i/2}{\th_\a-u_j-i/2}\,-
\sum_k^{J_v}i\log\frac{\th_\a-v_k+i/2}{\th_\a-v_k-i/2}\,, \label{DBAE1} \\
2\pi n_j^{u}&=&\sum_\b^L i \log\frac{u_j-\th_\b-i/2}{u_j-\th_\b+i/2}
+\sum_{i\neq j}^{J_u} i\log\frac{u_j-u_i+i}{u_j-u_i-i}\,,  \label{DBAE2}\\
2\pi n_j^{v}&=&\sum_\b^Li\log \frac{v_k-\th_\b-i/2}{v_k-\th_\b+i/2} +\sum_{l\neq k}^{J_v}i\log
\frac{v_k-v_l+i}{v_k-v_l-i}\,, \label{DBAE3}
\end{eqnarray}
where $u$'s and $v$'s are the Bethe roots appearing from the
diagonalization of (\ref{MBAE}) and characterizing each quantum
state. A quantum state with no such roots corresponds to the highest
weight  ferromagnetic state where all  spins of both kinds are  up.
Adding a $u$ ($v$) roots corresponds  to flipping one of the right
(left) $SU(2)$ spins, thus creating a magnon\footnote{This is
particularly clear from equations (\ref{DBAE2},\ref{DBAE3}) which in
the limit $\lambda\to 0$, when $\theta_\alpha\simeq 0$, are
precisely the usual Bethe equations for the diagonalization of an
Heisenberg hamiltonian for the periodic chain of length $L$,
originally soved by Hans Bethe \cite{BETHE}, provided we identify
the momentum of magnons with \beq e^{ip}=\frac{u+i/2}{u-i/2} \,.
\eeq}. The left and right charges of the wave function, associated
with the two $SU(2)$ spins are  given by
\beq  Q_L=L-2J_u\, , \qquad   Q_R=L-2J_v\,.  \la{QQ}\eeq

This  model with massive relativistic particles and the
asymptotically free UV behavior cannot look like  a consistent
quantum string theory. Only in the classical limit we can view it as
a string toy model obeying the  classical  conformal symmetry.
   In the classical case it is also easy to impose the Virasoro
   conditions. In the quasi-classical limit , we still can try to
   impose the Virasoro conditions as
   some natural constraints on  the quantum states. We will discuss this point latter.

\subsection{Quasi-classical limit} \label{quasi}

In  the classical limit the physical mass of the particle
\footnote{
For the $O(N)$ sigma model the beta function for the coupling is
given by $\beta\equiv \frac{\partial}{\partial \log\Lambda}\,
\sqrt{\lambda(\Lambda)}=N-2$ where $\Lambda$ is the cutoff of the
theory. The dynamically generated mass must be of the form
$m=\Lambda\,f(\sqrt{\lambda}\,)$. The functional form of $f$ is
fixed by the $\beta$ function upon imposing independence on the
cutoff of this physical quantity. Thus, for general $N$,
$-\log\mu=\frac{\sqrt{\lambda}}{N-2}+\O(1)$.}
\beq
 \frac{\mu}{2\pi} \sim  e^{-\sqrt{\lambda}/2}\,,
\eeq where $\lambda$ is the physical coupling at the scale $2\pi$,
vanishes since $\lambda\rightarrow \infty$. Moreover we should focus
on quantum states with large quantum numbers, i.e. we shall consider
a large number $L\to\infty$ of particles on the ring.

Let us now think of (\ref{DBAE1}-\ref{DBAE3}) as of the equations
for the equilibrium condition for a system of three kinds of
particles: ($\theta_\a$, $u_j$ and $v_k$), interacting between
themselves and experiencing the  external constant forces ($2\pi
m_\a$, $2 \pi n_j^{u}$ and $2 \pi n_k^{v}$). The particles of the
$\th$ kind  are also placed into the external confining potential
\beq V(z)=\mu\cosh(\pi Mz) \, \, , \, \,\qquad  z=\theta/M\la{BOX} \eeq
where
\begin{eqnarray}
\la{defM}M\equiv -\frac{\log\mu}{2\pi}\simeq\frac{\sqrt\lambda}{4\pi} \,.
\end{eqnarray}
\begin{figure}[t]\centering
        \includegraphics[scale=1]{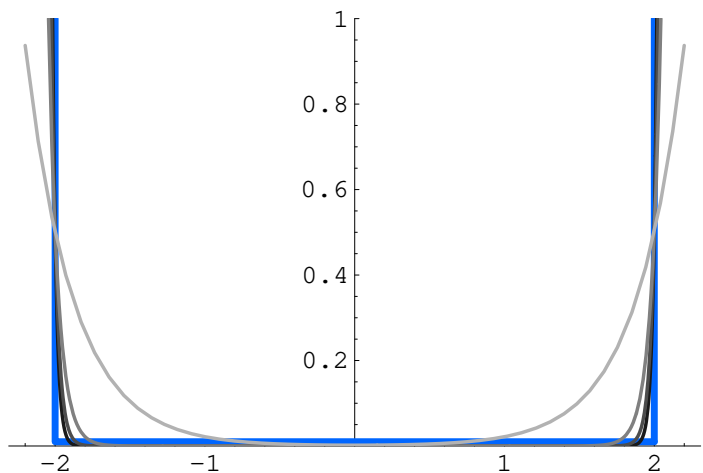}
            \caption{We plot $V(z)$ for $M=1,5,9,13$ (lighter to darker gray).
            It is clear that the potential approaches the blue
             box potential as
            $M\rightarrow
            \infty$.}
    \label{fig:box}
\end{figure}
In the classical limit the potential becomes a square box potential
with the infinite walls at $z=\pm2$  (see fig.\ref{fig:box}).
Moreover, since this is a large box for the original variables  we
can use the asymptotics (\ref{ASSS}) for the force between particles
of the $\theta$ (or $z$) type. The box potential provides the
appropriate boundary conditions for the density of  particles
interacting by the Coulomb force. Since they repeal each other the
density should be peaked around $z=\pm 2$. To find the correct
asymptotics close to these two points,  we can consider \eq{DBAE1}
as the equilibrium condition for the gas of Coulomb particles in the
box.

If the right and left modes (magnons) are not excited we have only
the states with  $U(1)$ modes. In the  classical limit, using the
Coulomb  approximation \eq{ASSS}, we have for this sector the
following Bethe equation
\begin{eqnarray}
\nn\mu \sinh\pi M z_\a-2\pi\,m&=& -\frac{1}{M}\sum_{\beta\neq
\alpha}^L\frac{1}{z_{\a}-z_{\b}}\,. \label{BAEnoUV}
\end{eqnarray}
In the continuous limit, the equation for the asymptotic density,
$L\sim M\rightarrow \infty$, is given, through the resolvent
$G_\th(z)=\frac{1}{M}\sum_{\beta=1}^L\frac{1}{z-z_{\b}}$ by
\beq\label{THETAG} \sG_\theta(z)=-2\pi m,\;\;\;\;\;\, z\in
\mathcal{C}_\theta\,, \eeq with inverse square root boundary
conditions near $\pm 2$. The analytical function $G_\theta(x)$
having a real part on the cut defined by \eq{THETAG}, with support
$[-2,2]$, with inverse square root boundary conditions (the only
compatible with the asymptotics at $z\to\infty$: $G_\theta(z)\to
\frac{L}{M}\frac{1}{z}$, is completely fixed:
\beq\label{RESTH}   G_\theta(z)=\(\frac{2\pi m\ z
+\frac{L}{M}}{\sqrt{z^2-4}}-2\pi m\), \qquad L>4\pi |m| M    \eeq
which gives for the density
\beq\label{DENTH}   \rho_\theta(z)=\frac{1}{\pi }\(\frac{2\pi m\ z
+\frac{L}{M}}{\sqrt{4-z^2}}\)  \,. \eeq
For a general solution with $u$ and $v$ magnons we will also
find the same asymptotics 
 \beq \rho(z)\equiv
\frac{1}{M}\sum_{\alpha=1}^{L}\delta\(z-z_\alpha\)\simeq \frac{2
\kappa_\pm}{\sqrt{2 \mp z}},\qquad z\to \pm 2. \la{rhowall} \eeq
with $\kappa_\pm$  yet to be  determined through the energy and
momentum of the solution, as we shall explain in the next section.

 We will be considering the scenario
where we have the same mode number $m_\alpha=m$ for all $z$
particles. As proposed in \cite{GKSV,Mann:2005ab} this is the
adequate set of states which will obey the Virasoro constraints in
the classical limit.

First, we will relate the $z$ behavior close to the walls,
characterized by the constants $\kappa_\pm$ with the energy and
momentum $E,P$ of the quantum state, as given by
(\ref{P_SSL},\ref{E_SSL}). Then we shall eliminate the $\theta$'s
from the system of Bethe equations by explicitly solving the first
one in the considered limit. Finally, we will justify  why we take
the same mode number $m$ for all $\theta$'s  by identifying the
longitudinal modes to the excited mode numbers $m_i$  in the Bethe
ansatz setup. This constraint on the states will correspond to the
Virasoro conditions, at least in the classical limit.

\subsubsection{Energy and momentum}
The total momentum can be calculated exactly, before any classical
limit\footnote{For the closed string theory we should take $P=0$ which gives the
level matching condition. Moreover, as we shall explain latter, we should also pick
the same mode number for all particles, $m_\alpha=m$. For the perturbative super SYM
applications
one should moreover take $S^u_p=0$ \cite{Minahan:2005jq}. Then we have the
well known formula $\sum_p n_p S_p^v=m L$ (see \cite{Kazakov:2004qf}
for details). }
\begin{eqnarray}\la{P_SSL}
P=\frac{\mu}{2\pi}\sum_{\a}\sinh(\pi\theta_\alpha)=m_p L_p-\sum_p n_p
S_p^u-\sum_p n_p S_p^v
\end{eqnarray}
where $L_p,S_p^u,\;S_p^v$ are the filling fractions, or the numbers
of Bethe roots with a given mode numbers $m_p,n_{u,p},n_{v,p}$.  To
prove this, it suffices to sum the \eq{DBAE1} for all roots
$\theta_\alpha$. The contribution of $S_0(\theta)$ terms cancels due
to antisymmetry while the second and third sums in the r.h.s. of
(\ref{DBAE1}) are replaced using  (\ref{DBAE2}) and (\ref{DBAE3}),
respectively.

Let us show how to calculate the energy (\ref{E_SSL}) which is a
fare less trivial task \cite{GKSV}. As a byproduct we will also
reproduce the total momentum from the behavior at  the singularities
at $z=\pm 2$ described by the residua $\kappa_\pm$. We want to
compute the sum
\beq \nn E\equiv \frac{\mu}{2\pi}\sum_\alpha
\cosh(\pi\theta_\alpha)\,, \eeq
but we  \textit{cannot} simply replace this sum by an integral and
use the asymptotic density $\rho_\th(z)$ to compute the energy. That
is because the main contribution to the energy comes from large
$\theta$'s, near the walls, where the expression for the asymptotic
density is no longer accurate. It is natural for the classical limit
since the particles become effectively massless and the
contributions of right and left modes are clearly distinguishable
and  located far from $\theta=0$. We notice that the energy is
dominated by large $\th$'s where, with exponential precision, we can
replace $\cosh \pi\theta_\a$ by $\pm \sinh \pi \th_\alpha$ for
positive (negative) $\theta_\a$. Furthermore, the contribution from
the $\th$'s in the middle of the box is also exponentially
suppressed since $\mu$ is very small. Thus we can pick a point $a$
somewhere in the box not too close to the walls. One can think of
$a$ as being somewhere  in the middle. Then,
\beq \la{ener} E= \sum_{z_\alpha>a} \frac{\mu}{2\pi} \sinh\( \pi
z_\alpha M\)- \sum_{z_\alpha<a}\frac{ \mu}{2\pi}\sinh \(\pi
z_\alpha M\)\, , \nn \eeq
 where, let us stress, the result is \textit{correct
independently of the point $a$ within the interval $-2<a<2$ with the
exponential precision}. Each sum of $\sinh \pi \theta_\a$  can be
substituted    by the corresponding r.h.s. of the  Bethe equation
(\ref{DBAE1}), thus giving
\beqa E&\simeq& \frac{i}{\pi}\!\!\!\!
\sum_{z_\b<a<z_\a}\!\!\!\!\log
S_0^2\(M\[z_\alpha-z_\beta\]\)
+\sum_{\alpha}m\;{\rm sign}(z_\alpha-a) \la{sum} \, \\
&-&\frac{1}{2\pi}\sum_{j,\alpha}  {\rm
sign}(z_\alpha-a)i \log\frac{Mz_\a-u_j+i/2}{Mz_\a-u_j-i/2} -\frac{1}{2\pi}\sum_{k,\alpha}  {\rm
sign}(z_\alpha-a)  i\log\frac{Mz_\a-v_k+i/2}{Mz_\a-v_k-i/2} \nn \eeqa
As mentioned above we assume all $m_\alpha$ to be the same
\footnote{as we will show it is this choice of states which
reproduces the finite gap solution of \cite{Kazakov:2004qf} we
mentioned in the first section. We will come back to this point at a
latter stage}. Now we can safely go to the continuous limit since in
the first term the distances between $z$'s are now mostly of the
order  $1$\footnote{ Moreover, it is very important that the
contribution from  $z$'s near the walls $\pm 2$ is now suppressed
since \eq{ASSS}
\begin{eqnarray*}
|\log S_0^2(M(2-z_\b))|>|\log S_0^2(M(2-a))|\sim 1/M.
\end{eqnarray*}}.
This allows to rewrite the energy, with $1/M$ precision, as follows
\beqa
E&\simeq& -\frac{M}{\pi}\int_{-2}^a dz\int_{a}^2 dw \frac{\rho_\theta(z)\rho_\theta(w)}{z-w}-\frac{M}{2\pi} \int \frac{\rho_\theta(z)\rho_u(w)}{z-w}{\rm sign}(z-a)\,dz\,dw \nn\\
&-&\frac{M}{2\pi} \int \frac{\rho_\theta(z)\rho_v(w)}{z-w}{\rm sign}(z-a)\,dz\,dw+m\,M  \int
\rho_\theta(z) {\rm sign}(z-a)\,dz\label{startingpoint} \eeqa
where we are now free to use the asymptotic density $\rho_\th(z)$.
By the use of Bethe equations, we managed to transform the original
sum over $\cosh$'s, highly peaked at the walls, into a much smoother
sum where the main contribution is now softly distributed along the
bulk and where the continuous limit does not look suspicious. From
the previous discussion we know that this expression does not depend
on $a$ provided $a$ is not too close to the walls. In fact, we can
easily see that it does not depend on $a$ \textit{at all} after
taking the continuous limit leading to the perfect box-like
potential. To prove it one notices that due to Bethe equations
\eq{DBAE1} the $a$-derivative of \eq{startingpoint} is zero for all
$a\in ]-2,2[$. Hence we can even send $a$ close to a wall:
$a=-2+\epsilon$, where $\epsilon$ is very small. But then the last
three terms in (\ref{startingpoint}) are precisely the momentum
(\ref{P_SSL}), as explained in the beginning of this section. To
compute the first
 term we can now use the asymptotics (\ref{ASSS},\ref{rhowall}).
 The contribution of this term
 is then given by
\beq \nn-\frac{M}{\pi}\int_{-2}^{-2+\epsilon} dz\int_{-2+\epsilon}^2
dw \frac{\rho_\theta(z)\rho_\theta(w)}{z-w}\simeq
-\int_{-2}^{-2+\epsilon} dz\int_{-2+\epsilon}^2 dw \frac{4 M
\kappa_-^2}{\pi(z-w)\sqrt{2+z}\sqrt{2+w}}\simeq  2\pi M \kappa_-^2
\eeq so that \beqa E\simeq 2 M \kappa_-^2 \pi+P\,. \la{energy} \eeqa
If we compute the $a$-independent integral (\ref{startingpoint})
near the other wall, i.e. for $a=2-\epsilon$, we find
\begin{eqnarray*}
E\simeq 2 M \kappa_+^2 \pi-P \,. 
\end{eqnarray*}
Therefore, equating the results one obtains the desired expressions
for the energy and momentum
\begin{eqnarray}
E\pm P=2\pi\,M\,\kappa_\pm^2 \la{EP}
\end{eqnarray}
through the singularities of the density of rapidities at $z=\pm 2$,
described by $\kappa_\pm$. Together with (\ref{defM}) this is
precisely the classical formula (\ref{EPclassic}).

\subsubsection{Elimination of $\theta$'s and AFS equations}
It is useful for what follows, to introduce some new notations. Using the Zhukovsky map
\beqa
z=x(z)+\frac{1}{x(z)} \,\, , \,\, |x(z)|>1
\eeqa
we define
\beqa
&&\nn y_j^{\,\pm}\equiv x\(\frac{u_j\pm \, i/2}{M}\),\;\;\;\;\; y_j\equiv x\(\frac{u_j}{M}\)
\eeqa
with the similar expressions for $v_l$ given by $\tilde y_l^\pm$ and $\tilde y_l$.

In this section, for the purposes of comparison with the asymptotic
AFS Bethe ansatz for the N=4 SYM theory,  let drop the $v$ magnons,
$J_v=0$. Their contributions will be easily restored later. As
explained at the beginning of this section we can write the first
Bethe equation, (\ref{DBAE1}) as \beqa
 \pint_{-2}^2\frac{\rho(w)}{z-w}dw=-\sum_j^{J_u}i\log\frac{M z-u_j+i/2}{M z-u_j-i/2}
 -2\pi m,\,\,\,\, z\in
[-2,2]\,. \nn
 \eeqa
The solution to this Riemann-Hilbert problem with the  boundary
conditions and the normalization given by (\ref{rhowall}) looks as
follows \cite{Gromov:2006cq}
 \beqa\label{FTTRM} \rho(z)&=&\frac{1}{\pi
\sqrt{4-z^2}}
\[\(2\pi m + i \sum\limits_{j=1}^{J_u} \log\frac{y_j^-}{y_j^+}\)\,z
+ \frac{L}{M}+
2i\sum\limits_{j=1}^{J_u}\(\frac{1}{y_j^+}-\frac{1}{y_j^-}\) \] \nn
\\&-&\frac{1}{\pi}\sum\limits_{j=1}^{J_u} \log\(\frac{x(z) y_j^+-1}{x(z)
y_j^--1}\,\,\,\,\frac{x(z)-y_j^-}{x(z)-y_j^+}\)\ . \eeqa
We want to focus on such states  that the momentum $P$ related to
the asymptotics close to the walls by (\ref{EP}), vanishes. Thus we
should set to zero the first  term in the r.h.s. of \eq{FTTRM}:
\beq P= m-\frac{i}{2 \pi} \sum\limits_{j=1}^{J_u}
\log\frac{y_j^+}{y_j^-}=0 \,. \eeq
Then, plugging this  density into (\ref{DBAE2}), integrating over
the rapidities and exponentiating the result, we find
\cite{Gromov:2006cq}
\beq \la{AFS}\(\frac{y_k^+}{y_k^-}\)^L=
\prod_{j\neq k}^{J_u} \frac{u_k-u_j+i}{u_k-u_j-i}
\,\sigma^2(u_j,u_k)\,,
\eeq
where the
 ``dressing" factor
$\sigma^2$ is given by
\beq \la{sigma2}\sigma^2(u_j,u_k)=
\(\frac{1-1/(y_j^-y_k^+)}{1-1/(y_j^+ y_k^-)}\)^{-2} \(\frac{y_j^-
y_k^--1}{y_j^-y_k^+-1} \frac{y_j^+ y_k^+-1}{y_j^+
y_k^--1}\)^{2i(u_j-u_k)}\,. \eeq These are precisely the AFS
equations conjectured in \cite{Arutyunov:2004vx} as the asymptotic
Bethe ansatz equation for the $SU(2)$ sector of $N=4$ SYM theory
\footnote{A similar derivation of the BDS equation in  N=4 SYM
theory was given in \cite{Rej:2005qt} starting from the Hubbard
model}. The dispersion relation for these dressed magnons can be
read off from the asympotics of the density \eq{FTTRM} close to the
walls \footnote{In the context of the AdS/CFT correspondence
$\kappa=\kappa_-=\kappa_+$ is the energy with respect to the AdS
global time $Y$ equal to the dimension of the corresponding SYM
operator, see (\ref{Y}).} \beq \Delta\equiv \sqrt{\lambda} \,\kappa=
L +2 Mi\sum\limits_{j=1}^{J_u} \(\frac{1}{y_j^+}-\frac{1}{y_j^-}\)
\,. \la{D} \eeq

\subsubsection{Classical limit and  KMMZ algebraic curve}

To consider the classical limit we trivially restore the $v$ roots
from the previous calculation, to find
\beqa
\(\frac{y_k^+}{y_k^-}\)^L&=&\prod_{j\neq k}^{J_u}
\frac{u_k-u_j+i}{u_k-u_j-i} \,\sigma^2(u_j,u_k)\,  \prod_{l=1}^{J_v}
\sigma^2(v_l,u_k) \,, \eeqa
and similarly for $\tilde y_k$, and consider the limit where
$J_u,J_v,L\sim M$, so that the $u$ and $v$ roots also scale as $M$.
Then the expansion of this equation, after taking the $\log$'s,
gives to the leading order in $1/M$
 \beqa \la{baecl}\pi
n_k&=&\frac{\frac{L}{2M}y_k+2\pi m}{1-y_k^2}
+\frac{1}{y_k^2-1}\frac{1}{M}\sum_{l=1}^{J_v}\frac{1}{1/y_k-\tilde
y_l} +\frac{y^2_k}{y^2_k-1}\frac{1}{M}\sum_{j\neq
k}^{J_u}\frac{1}{y_k-y_j} \,. \eeqa
 Finally we can define the quasimomentum
\cite{Gromov:2006cq}
\beq p(x)=\frac{\frac{L}{2M}x+2\pi m}{1-x^2}
+\frac{1}{x^2-1}\frac{1}{M}\sum_{j=1}^{J_v}\frac{1}{1/x-\tilde y_j}
+\frac{x^2}{x^2-1}\frac{1}{M}\sum_{j=1}^{J_u}\frac{1}{x-y_j} \,.
\la{disp}
\eeq
Let us explain how it becomes precisely the quasimomentum we had in
the context of the algebraic curve in section \ref{first} in the
classical theory. It is clear that we indeed have the asymptotics
(\ref{PXZERO},\ref{GLCH}) close to $x=0,\infty$. Then, to relate the
residues of \eq{disp} to the ones found from the algebraic curve in
\eq{LCH}, we expand (\ref{D}) in our limit as follows: \beq
\Delta=L+\sum_j\frac{2}{y_j^2-1}+\sum_l\frac{2}{\tilde y_l^2-1} \eeq
and check that this is indeed what one finds from the quasimomenta
we just defined. Finally, when we consider a large number of magnons
$J_u,J_v$ the roots in \eq{disp} condense into a number of one
dimensional supports, the sums becoming the integrals along these
lines giving  the same square root cuts as we had in the finite gap
construction.

\subsubsection{Geometric proof}

\begin{figure}[t]\centering
        \includegraphics[scale=0.7]{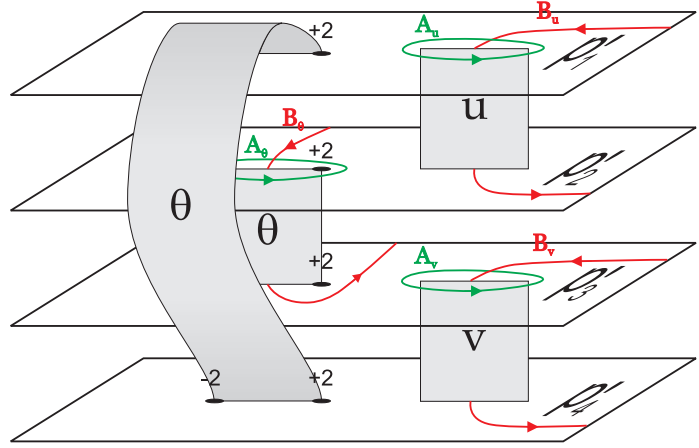}
    \caption{Structure of the curve coming from the Bethe ansatz side.  This figure is related with fig.1 by means of the Zhukovsky map.}
            \label{fig:sheets}
\end{figure}

The roots solving (\ref{DBAE1},\ref{DBAE2},\ref{DBAE3}) with the
same mode number will condense into a single square  root cut. When
we consider more than one type of mode numbers we see that the
particles condense into a few distinct supports, one for each
distinct mode number
$$
\mathcal{C}= \mathcal{C}_1 \cup \dots \cup \mathcal{C}_K \,.
$$
We can now rescale the Bethe roots \beq (u,v,\theta)=M(x,y,z)
\la{scale} \eeq and define
\begin{eqnarray}
p_1=-p_2&=&\frac{1}{M}\sum_{i=1}^{J_u}
\frac{1}{z-x_i}- \frac{1}{2M}\sum_{\beta=1}^{L}\frac{1}{z-z_\beta} \nn \\
p_3=-p_4&=&\frac{1}{M}\sum_{l=1}^{J_v}
\frac{1}{z-y_l}- \frac{1}{2M}\sum_{\beta=1}^{L}\frac{1}{z-z_\beta}\,.
\la{p1p2p3p4}
\end{eqnarray}
Then we can recast the Bethe equations in this scaling limit as
follows \beqa
x\in {\cal C}_u,&&\;\;\;\;\;{p_1}^+-{p_2}^-=2\pi  n_u\nn\\
x\in {\cal C}_\theta,&&\;\;\;\;\;{p_2}^+-{p_3}^-=2\pi m \la{CONTBAE}\\
x\in {\cal C}_v,&&\;\;\;\;\;{p_3}^+-{p_4}^-=2\pi  n_v \nn\\
x\in {\cal C}_\theta,&&\;\;\;\;\;{p_4}^+-{p_1}^-=2\pi m,  \nn \eeqa where
we
\begin{itemize}
\item{considered, as in the preceding section, one single mode number
$m$  for all rapidities;}
\item{dropped the momentum $\mu \sinh \theta$. As we explained in section \ref{quasi}
we can do this provided we replace it by the boundary conditions
(\ref{rhowall}).}
\end{itemize}
These equations tell us that $p'_1(z), p'_2(z),p'_3(z), p'_4(z)$ form
four sheets of the Riemann surface of an analytical function $p'(z)$
(see fig.\ref{fig:sheets}).

They can also be written as holomorphic integrals around the
infinite B-cycles: \beqa \oint_{B^u_{j}}dp &=& 2\pi
n_{u,j}\qquad n_j=1,\ldots,K_u\nn\\
\oint_{B^v_{j}}dp &=& 2\pi
n_{v,j}\qquad n_j=1,\ldots,K_v\la{BCYCLES}\\
\oint_{B^\theta}dp &=& 2\pi m\nn \eeqa
where the the first two conditions correspond to the  equations in the
first and third line of (\ref{CONTBAE}), respectively, while the last
one corresponds to any of the equations of the second and fourth lines
of (\ref{CONTBAE}). The $B$ cycles are defined as in
fig.\ref{fig:sheets}.

We found two Riemann surfaces which we plotted in figures 1 and
\ref{fig:sheets}. The equivalence between these two curves is
achieved through the Zhukovsky map \cite{GKSV}
$$
z=x+\frac{1}{x}
$$
and amounts to the equivalence between the finite gap solutions for
the classical theory and the Bethe ansatz solutions in the scaling
limit.

\subsubsection{Virasoro modes}
We established the equivalence between
\begin{itemize}
\item{
all classical solutions following from the  PCF action
(\ref{ACTIONSU2}) and subject to the Virasoro conditions ${\rm tr}
\(j_\tau\pm j_\sigma\)^2=-2 \kappa_{\pm}^2$ as described by the
construction of the algebraic curve of section \ref{first}. }
\item{
and the Bethe ansatz quantum solution (\ref{DBAE1}-\ref{DBAE2}) in the scaling limit (\ref{scale}) with all rapidities $\theta_\alpha$ having the same mode number $m$.
}
\end{itemize}
In the context of string theory one is interested in quantizing the
Polyakov string action
 \begin{equation}
S=\frac{\sqrt{\l}}{8\pi} \int d\s d\tau \sqrt{h}\, h^{ab} \(\tr\,\d_a g^\dag\d_b g -\d_a Y \d_b Y \)  \,.
\end{equation}
Due to its local reparametrization and Weyl symmetries one can then
fix the target space time $Y$ as in (\ref{Y}) and reduce  the action
to (\ref{ACTIONSU2}). However, due to the residual reparametrization
symmetry \beq \tau \pm \sigma \rightarrow f_\pm(\tau \pm \sigma) \,,
\eeq one must keep in mind that the original presence of the
world-sheet metric field imposes that the stress energy tensor
vanishes. This is precisely the Virasoro conditions.

On the other hand, from the field theory point of view the Bethe ansatz equations (\ref{DBAE1}-\ref{DBAE3}) should describe all possible states of the theory, not only those for which
\beq
\langle \psi | T^{ab} | \phi \rangle =0 \,.
\eeq

Thus, in view of the equivalence we proved, we are lead to the
conclusion that if we start with some classical solution with one
$\theta$ cut and some $u$ and $v$ cuts, the excitation of additional
microscopic $\theta$ cuts should correspond to the inclusion of the
longitudinal modes which we drop in the context of string theory.
Indeed, these massless (from the world-sheet point of view)
excitations coming from our conformal gauge choice, appear if one
expands the action around the classical solution without fixing the
Virasoro conditions from the beginning (see for instance expression
2.7 and the discussion following it in \cite{Frolov:2003tu}).  In
this section we verify this claim therefore justifying this single
$\theta$ cut restriction, first proposed in \cite{Mann:2005ab} and
given the interpretation as the Virasoro condition in  \cite{GKSV}.

In (\ref{sum}) we computed the energy of a quantum state where  all
mode numbers $m_\alpha=m$ were the same. If we change the mode numbers
of a few $\theta$'s  we will have a macroscopic support with
particles having the mode number $m$ surrounded by some microscopic
domains, linear supports, with mode numbers $m_\beta<m $ (to the left of
it) and $m_\beta>m $ (to its right).

Let us assume  that we excite them one at a time  and focus on the
first particle whose mode number we change. Before we do it, it is
in equilibrium due to the exponential force exerted by the wall of
the box (\ref{BOX}) and by (an equal) force produced by all the
other particles and by the constant force $2\pi m$ -- see
(\ref{DBAE1}). When we change the particle mode number the constant
force increases pushing the particle against the wall. However since
the forces are exponential the shift will be very small, much
smaller than $1/M$ - the characteristic distance between the
neighboring  rapidities. Then let us consider the particles in the
middle of the box, the ones whose position is well described by the
asymptotic density $\rho(z)$. They only feel the change in mode
number through the new position of the corresponding $\theta$
particle. Since this shift is very small the asymptotic density, to
the order we are interested, is not changed. Thus, in this procedure
of changing a few mode numbers we conclude that, when going to the
continuous limit in (\ref{sum}), only the second term will lead to a
different result so that 
\beq \delta E=\sum_n |n| N_{m+n} 
\eeq 
where
$N_{n}$ is the number of particles with mode number $n$. We
found in this way the massless (world-sheet) modes associated with
the local reparametrization symmetry of the world-sheet. These modes
appear when considering the fluctuations around a classical solution
\cite{Frolov:2003tu} and are the only ones not taken into account by
the finite gap algebraic curve \cite{GV}.


\acknowledgments

We are grateful to Kazuhiro Sakai for collaboration in \cite{GKSV}. The work of V.K. was partially
supported by European Union under the RTN contracts MRTN-CT-2004-512194 and by INTAS-03-51-5460
grant. The work of N.G. was partially supported by French Government PhD fellowship, by
RSGSS-1124.2003.2 and by RFFI project grant 06-02-16786. P.~V. is supported by the Funda\c{c}\~ao
para a Ci\^encia e Tecnologia fellowship SFRH/BD/17959/2004/0WA9.

\end{document}